\documentclass[12pt]{article}
\usepackage{epsfig}

\newcommand{\nc}{\newcommand}
\nc{\be}{\begin{equation}}
\nc{\ee}{\end{equation}}

\begin{document}

\begin{center}

\vspace{0.5cm}

{\bf \Large Evaluation of Multi-Box Diagrams in Six Dimensions}
\vspace{1cm}

{\bf \large  D.I.Kazakov}\vspace{0.5cm}

{\it Bogoliubov Laboratory of Theoretical Physics,\\
Joint Institute for Nuclear Research, Dubna, Russia,\\[0.1cm] and\\
Institute for Experimental and Theoretical Physics, Moscow, Russia\\[0.1cm] and\\
Moscow Institute of Physics and Technology, Dolgoprudny, Russia
}\vspace{1cm}

\abstract{We present a simple method which simplifies the evaluation of the on-shell multiple box diagrams reducing them to triangle type ones. For the $L$-loop diagram one gets the expression in terms of Feynman parameters  with $2L$-fold integration. As examples we consider the 2 and 3 loops cases, the numerical integration up to six loops is also presented. The method is valid in six dimensions where neither UV not IR divergences appear.}
\end{center}

\section{Introduction} 
The box-type diagrams being difficult for evaluation have attracted much attention in recent years due to the appearance in multiloop calculations of amplitudes in SUSY theories~\cite{Dixon}. It was shown that  in maximal SYM theories  when calculating the amplitudes of many-particle scattering the bubble and triangle diagrams cancel and only the boxes survive~\cite{Bern}. At the same time, long ago the ladder diagrams in four dimensions  were calculated in an arbitrary order of PT, albeight off-shell~\cite{UsD,UsB}. When going on-shell they contain infrared divergences and the proper $\varepsilon$-expansion was constructed in 2 and 3 loops~\cite{Smir}. Since the form of the integrand of the four point amplitude in maximal SYM theories is fixed by dual conformal invariance and essentially the same in any dimension, the same boxes appear in the other dimensions~\cite{Arc}

Recently, we considered the amplitudes in the D=6 N=(1,1) SYM theory and encountered the same box diagrams but in six dimensions~\cite{BKV}. They are UV and IR convergent and  at lower orders may in principle be extracted from the known Mellin-Barnes representations~\cite{SmirBook}. However, the remaining MB integrals are complicated starting from 2 loops. 

Here we present the method which is applicable in six dimensions and allows us to reproduce the known results and 
 proceed to higher orders of PT in a straightforward way.

\section{Reduction of the box diagrams to triangles}
\begin{figure}[htb]
\begin{center}
\leavevmode
\includegraphics[height=1.5cm]{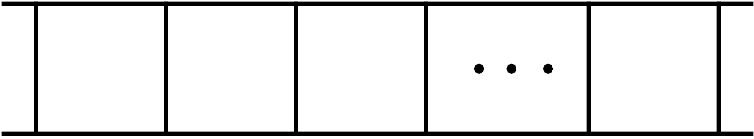}
\end{center}
\caption{The ladder type diagram}
\label{ladder}
\end{figure}

Consider the $L$-loop ladder diagram shown in Fig.\ref{ladder}  in the massless theory in six dimensions.  In  the momentum space it depends on 4 external momenta and integration is taken over all loop momenta, while in the coordinate space it depends on 4 external coordinates and integration is taken over all internal vertices. 
These two expressions are related via the Fourier transform. One should have in mind that the propagator in the coordinate space depends on the space-time dimension
\begin{equation}
\int d^Dp \frac{e^{ipx}}{p^2}=i \pi^{D/2}2^{D-2}\frac{\Gamma(D/2-1)}{\Gamma(1)}\frac{1}{(x^2)^{D/2-1}},
\label{four}
\end{equation}
and in six dimensions one has $1/(x^2)^2$.

\begin{figure}[h]
\begin{center}
\leavevmode
\includegraphics[height=3.5cm]{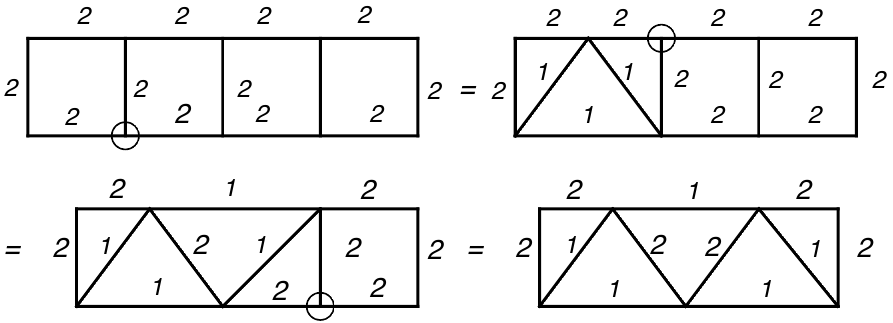}
\end{center}
\caption{Reduction of boxes. The indices of the lines in the coordinate space are shown}
\label{reduce}
\end{figure}

Now let us start from the coordinate space. One can notice that in all internal vertices the sum of the indices of the lines (index of a line is the power in the propagator) is equal to 6=2+2+2, i.e. the value of the space-time dimension. Such a vertex is called unique~\cite{Uniq} and obeys the famous uniqueness or star-triangle relation
\be
\includegraphics[height=1.5cm]{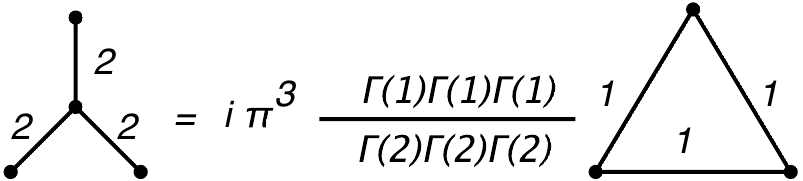},\label{star}
\ee
where in the coordinate space one integrates over the vertex and has no integration in the r.h.s. Thus, all the vertices in Fig.\ref{ladder} are unique, and one can easily reduce the ladder diagram to the sequence of triangles. The procedure consists of consequent use of eq.(\ref{star}) (see Fig.\ref{reduce}).

\begin{figure}[h]
\begin{center}
\leavevmode
\includegraphics[height=2cm]{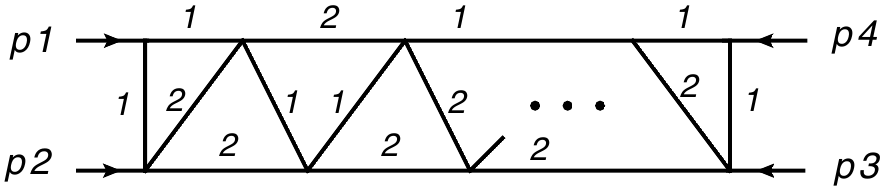}
\end{center}
\caption{The resulting ladder diagram in the momentum space. The indices of the lines in the momentum space are shown}
\label{momlad}
\end{figure}

The next step is to go to the momentum space. The diagram in the momentum space has the same topology but the indices of the lines have to be replaced by the dual ones ($2\to 1, 1\to 2$), and one has to integrate over the loops.
Performing this procedure one gets the diagram shown in Fig.\ref{momlad} (we omit the factors of $i, \pi$ and $2$, they are canceled eventually, and one is left with the  obvious loop factors). 

This is the diagram we are going to calculate.  It looks as if it contains an additional loop, however, as we will see, eventually one has $L-1$ loops to evaluate.  We will be interested in the on-shell value. This means that
all external momenta obey $p_i^2=0$. This simplifies the calculation and the diagram appears to be the function of
 two variables: the usual Mandelstam  $s=(p_1+p_2)^2=(p_3+p_4)^2$ and $t=(p_1+p_4)^2=(p_2+p_3)^2$.

\section{The two-loop example}

\begin{figure}[h]
\begin{center}
\leavevmode
\includegraphics[height=2.4cm]{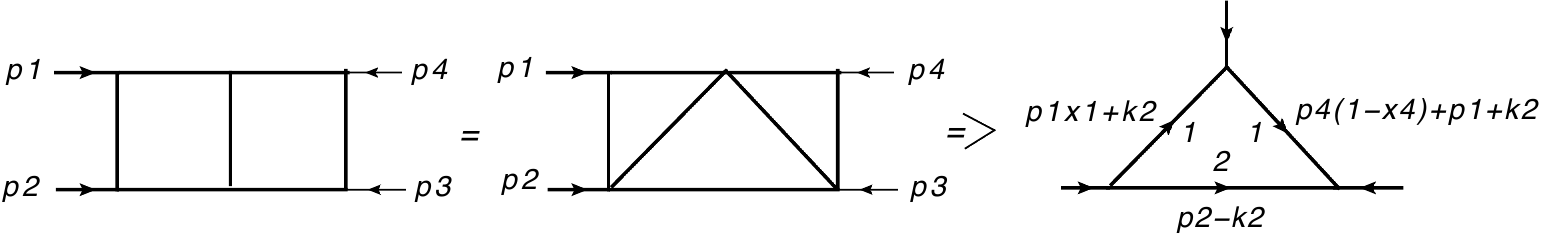}
\end{center}
\caption{The resulting two loop diagram}
\label{two}
\end{figure}

Before proceeding to the two loop case we present one simple observation. The triangles that are situated at the 
edges of the diagram can be easily evaluated, and the result is useful to present in the form
\be
\int \frac{d^6k}{k_1^2(p_1+k_1)^2(k_2-k_1)^2}=i(-\pi)^3\int_0^1 dx_1 \frac{1}{(p_1x_1+k_2)^2},  \hspace{1cm} \mbox{for} \ \ \ \ p_1^2=0. \label{tr}
\ee
This way the loops at the edges are simplified and one actually has the $L-1$ loop diagram. In the two loop case
one gets the diagram shown in Fig.\ref{two}.
Thus, for the double box diagram one has
\be
DBox(s,t)=\int_0^1 dx_1 dx_4 \int \frac{d^6k_2}{(p_1x_1+k_2)^2(p_2-k_2)^4(p_4(1-x_4)+p1+k2)^2}.
\ee
This integral can be evaluated by introducing Feynman parameters, which gives
\be
DBox(s,t)=\int_0^1 \frac{dx_1dx_2dx_3dx_4(1-x_2)}{s(x_1(1-x_3)+x_2x_3)(1-x_2)+t(1-x_4)x_2x_3(1-x_3)(1-x_1)}.
\ee
To deal with the integrals over Feynman parameters, it is useful to apply the Mellin-Barnes transformation.
We use twice the following relation:
\be
\frac{1}{(X+Y)^\lambda}=\frac{1}{\Gamma(\lambda)}\frac{1}{(2\pi i}\int_{-i\infty}^{i\infty} dz
\Gamma(\lambda+z)\Gamma(-z)\frac{X^z}{Y^{\lambda+z}},
\ee
and  then evaluate the parametric integrals. This gives
\begin{eqnarray}
DBox(s,t)&=&\frac{1}{s}\frac{1}{(2\pi i)^2}\int_{-i\infty}^{i\infty}dz_1dz_2\left(\frac ts\right)^{z_1}\Gamma(-z_1)\Gamma(-z_2)\Gamma(-1-z_1-z_2)\Gamma^2(1+z_1)\nonumber \\ 
&&\hspace{5cm}  \Gamma(1-z_1)\Gamma(1+z_1+z_2)\Gamma(z_2).
\end{eqnarray}
The integral over $z_2$ can be evaluated by using the first Barnes lemma\cite{SmirBook}, and one finally arrives at
\be
DBox(s,t)=\frac{1}{s}\frac{1}{2\pi i}\int_{-i\infty}^{i\infty}dz_1\left(\frac ts\right)^{z_1}\Gamma^3(-z_1)\Gamma^3(1+z_1)\frac{2z_1}{1+z_1}\left(\psi(1+z_1)-\psi(1)\right).\label{db}
\ee
The remaining integral is reduced to the sum of residues at integer points and reproduces the known answer~\cite{Tausk,BKV}.  Notice that our way to arrive at (\ref{db}) is much simpler than the usual MB evaluation which requires 4-fold MB integration~\cite{SmirBook,BKV}.

\section{The three-loops example for t=0}
Consider now the three loop case. Since the answer is too complicated and contains special functions, we constrain ourselves to the  limit when $t\to 0$.  It should be mentioned that this limit exists and the ladder diagram in this limit behaves as
\be
 Box^n(s,t)\to \frac 1s \left(c_n+O(\frac ts)\right).\label{coef}
\ee
In particular, from eq.(\ref{db}) it follows that $c_2=\pi^2/3$. Our aim here is to calculate $c_3$. The standard MB representation leads to 7-fold integration~\cite{SmirBook}.

Following the procedure described above we have the following 2-loop integral:
\be
TBox(s,t=0)=\int_0^1 dx_1 dx_6\int \frac{d^6k_2d^6k_3}{(p_1x_1+k_2)^2(p_2-k_2)^4(k_3-k_2)^2(p_1+k_3)^4(p_2x_6-k_3)^2}.\label{TB}
\ee
Using Feynman parameters (2 for each loop) one can reduce eq.(\ref{TB}) to
\be
TBox(s,t=0)=\frac 1s \int_0^1 dx_1\cdots dx_6\frac{x_2(1-x_3)(1-x_4)}{Poly_3},
\ee
where
\be
Poly_3=x_1x_2x_3(1-x_3)x_5+(x_1x_3x_4x_5+1-x_4)[x_6(1-x_5)+x_5(1-x_3)](1-x_2).
\ee
Now one can use the multiple MB transformation to get
\begin{eqnarray}
TBox&=&-\frac 1s\frac{1}{(2\pi i)^3}\int_{-i\infty}^{i\infty}dz_1dz_2dz_3\frac{\Gamma^2(-z_1)(1+z_1)}{\Gamma(2-z_1)}\Gamma(1+z_1+z_2)\Gamma(1+z_1+z_3)\nonumber \\ 
&&\frac{\Gamma^2(-z_2)\Gamma(-z_3)\Gamma(2+z_2)\Gamma(1+z_3)\Gamma(1-z_3)\Gamma(1-z_1-z_2-z_3)}{z_2(1+z_3)\Gamma(1-z_2-z_3)}.
\end{eqnarray}
The integral over $z_3$ can be evaluated by using the first Barnes lemma, and one gets (we change the sign of $z$)
\begin{eqnarray}
TBox&=&-\frac 1s\frac{1}{(2\pi i)^2}\int_{-i\infty}^{i\infty}dz_1dz_2\frac{(1-z_1)(1-z_2)}{z_1z_2}\Gamma(-1-z_1)\Gamma(z_1)\Gamma(-1-z_2)\Gamma(z_2)\nonumber\\
&&\hspace{-1cm}\Gamma(1+z_1+z_2)\Gamma(-z_1-z_2)\left[\psi(2)-\psi(2-z_1)+\psi(z_1+z_2)-\psi(z_2)\right].
\end{eqnarray}
The remaining 2-fold integral is more complicated but again can be taken by using Barnes lemmas. For this  purpose we used the dedicated computer code~\cite{Smirnov:2009up,Czakon:2005rk} reducing the integral to the standard form. The result of the integration is
\begin{eqnarray}
TBox&=&-\frac 1s\frac{1}{2\pi i}\int_{-i\infty}^{i\infty}dz_1\nonumber\\
&&\hspace{-2cm}1/(6 z_1^5 (-1 + z_1^2)) \pi^2 \csc^2(\pi z_1) (-6 + 
    z_1 (24 \gamma_E^2 z_1 (-1 + z_1^2) - 
       24 \gamma_E (1 + (-1 + z_1) z_1^2)\nonumber \\ &&\hspace{-2cm}+ 
       z_1 (6 (-2 + z_1) z_1 - 7 \pi^2 (-1 + z_1^2))) + 
    12 \pi z_1 ((-1 - 2 \gamma_E z_1 + 
          z_1^3) \cot(\pi z_1\nonumber)\\&&\hspace{-2cm} + \pi z_1 (-1 + 
          z_1^2) \csc^2(\pi z_1)) + 
    6 z_1 (4 (-1 - 2 \gamma_E z_1 + 
          z_1^3\nonumber\\&&\hspace{-2cm} + \pi z_1 (-1 + 
             z_1)) \psi(z_1) + 
       4 z_1^2) \psi^2(z_1) + (z_1 - z_1^3) \psi'(-1 + z_1)). \label{mbres}
 \end{eqnarray}
 The final integration is performed by taking the residues at integer points and summing up the obtained series.
 The latter can be done using the set of formulas from~\cite{SmirBook}. The result  for the desired coefficient is
 \be
 c_3=-\pi^2 + \frac{31 \pi^6}{1890} -8\zeta(3)+ 4 \zeta^2(3).\label{res}
 \ee
 This is a bit disappointing answer since after $c_1=\pi^2/2$ and $c_2=\pi^2/3$ one might expect a simple sequence
 at every loop.  With eq.(\ref{res}) this hope disappears and one is left with numerical answers.
      
\section{Numerical integration in higher loops}

Numerical values for the coefficients $c_n$ from (\ref{coef}) are summarized in Table 1. We complete this table by numerically evaluating the 4-, 5- and 6-loop box diagrams. 
\begin{table}[h]
\begin{center}
\begin{tabular}{|l|c|c|c|c|c|c|}
\hline
Loop & 1 & 2& 3 & 4 & 5 &6 \\ \hline
Value & $\pi^2/2$ & $\pi^2/3$ &  $-\pi^2 + 31 \pi^6/1890 -8\zeta(3)+ 4 \zeta^2(3)$ & & & \\ \hline
Numerics & 4.93 & 3.29 & 2.06 & 2.05& 2.42& 3.13\\
\hline
\end{tabular}
\caption{The values of the coefficients $c_n$}
\end{center}
\end{table}

To get numerical answers, we use the same procedure as above but stop at the level of Feynman parametrization without performing the MB transformation. Then in the $L$-loop order one has $2L$ parametric integrals from 0 to 1.
One starts from the left and carries out the 6-dimensional integration introducing 2 Feynman parameters for each loop.
The triangles on the left and right edges are taken using eq.(\ref{tr}). Each integration gives the integrand of a typical
form
$$\frac{1}{k_i^2+k_i q_i+m_i^2},$$
where $k_i$ is the momentum of the next integration, $q_i$ is some external momentum and $m_i^2$ is $s$ times some Feynman parameters. Thus, one has the triangle type integral again and can continue the same way.

We present below the resulting integrals for the 4-, 5- and 6-loop box diagrams.  They all have the same form
\be
c_L=\int_0^1 dx_1 \cdots dx_{2L} \frac{x_2(1-x_3)\cdots x_{2L-2}(1-x_{2L-1})}{Poly_L},
\ee
where the polynomials at the 4-, 5-, and 6-loop order are
\begin{eqnarray}
Poly_4&=&x_1 x_2 (1 - x_3) x_3 x_5 x_7 + (1 - x_2) (1 - x_3) x_4 x_5 (1 - x_5 + 
    x_1 x_3 x_5) x_7\nonumber\\ & +& (1 - x_2) (1 - x_4) (1 - 
    x_7 + (1 - x_3) x_5 x_7) ((1 - x_5 + x_1 x_3 x_5) x_6 x_7\nonumber\\ && + (1 - x_6) x_8),\\
Poly_5&=&x_1 x_2 (1 - x_3) x_3 x_5 x_7 x_9 + (1 - x_2) (1 - x_3) x_4 x_5 (1 - x_5 + 
    x_1 x_3 x_5) x_7 x_9\nonumber \\ &+& (1 - x_2) (1 - x_4) (1 - x_5 + x_1 x_3 x_5) x_6 x_7 (1 -
     x_7 + (1 - x_3) x_5 x_7) x_9 \nonumber\\&+& (1 - x_2) (1 - x_4) (1 - x_6) (1 - 
    x_9 + (1 - x_5 + x_1 x_3 x_5) x_7 x_9) (x_{10} (1 - x_8)\nonumber\\& +& (1 - 
       x_7 + (1 - x_3) x_5 x_7) x_8 x_9),\\
Poly_6&=&x_1  x_2 (1 - x_3) x_3 x_5 x_7 x_9 x_{11}+ 
 (1 - x_2) (1 - x_3) x_4 x_5 (1 - x_5 + x_1 x_3 x_5) x_7 x_9  x_{11}\nonumber\\&+& 
  (1 - x_2) (1 - x_4) (1 - x_5 + x_1 x_3 x_5) x_6 x_7 (1 - 
    x_7 + (1 - x_3) x_5 x_7) x_9  x_{11}\nonumber\\&+& 
  (1 - x_2) (1 - x_4) (1 - x_6) (1 - x_7 + (1 - x_3) x_5 x_7) x_8 (1 - 
    x_9) x_9  x_{11}\nonumber\\&+& 
 (1 - x_2) (1 - x_4) (1 - x_5 + x_1 x_3 x_5) (1 - x_6) x_7 (1 - 
    x_7 + (1 - x_3) x_5 x_7) x_8 x_9^2  x_{11}\nonumber\\&+& (1 - x_2) (1 - x_4) (1 - x_6) (1 - 
    x_8) (1 - x_{11} + x_{11} (1 - x_7 + (1 - x_3) x_5 x_7) x_9) \nonumber\\ &&((1 - x_{10}) x_{12} 
    x_{10} x_{11} (1 - x_9 + (1 - x_5 + x_1 x_3 x_5) x_7 x_9)).
\end{eqnarray}
Numerical evaluation of these integrals can be performed in the standard Mathematica or in more sophisticated 
codes using sector decomposition strategy~\cite{Bogner:2007cr,Hahn:2004fe}  or FIESTA~\cite{Fiesta}.

The results are
\be
c_4\approx 2.0520\pm 0.0001, \ \ \  c_5\approx 2.4257\pm 0.0008, \ \ \ c_6\approx 3.130\pm0.005 .\label{numres}
\ee
If needed this can be continued in higher loops.
\section{Conclusion}
One can see that reduction of boxes to triangles is very efficient and enables one to calculate the  higher order $L$-loop diagrams
with the price of the $2L$-fold parametric integration which can  be easily carried out numerically. Unfortunately, the method
is linked to six dimensions since only in this case one has unique vertices and can use the power of the star-triangle relation. 
\section*{Acknowledgments}
The author is grateful to V.Smirnov for running the code that helped to get eq.(\ref{mbres}) and for useful hints in evaluation of the integrals, and to A.Pikelner for the accurate check of numerical results (\ref{numres}) with the help of the sector decomposition  and FIESTA programs. I would also like to thank V.Belokurov and A.Grozin for their help in finding the right way and O.Tarasov, A.Kotikov, A.Isaev, K.Chetyrkin, L.Bork and D.Vlasenko for useful discussions. 
Financial support from RFBR grant \# 14-02-00494 is kindly acknowledged. Thanks are also to Mr. Mac who always kept me in good shape.


\begin{thebibliography}{90}
\bibitem{Dixon} Z.~Bern, L.~J.~Dixon, D.~A.~Kosower \emph{Progress in One-Loop QCD
Computations}, Ann.Rev.Nucl.Part.Sci. \textbf{46} (1996) 109,
arXiv:hep-ph/9602280 v1.
\\
Z.~Bern, L.~J.~Dixon, D.A.~Kosower \emph{On-Shell Methods in
Perturbative
QCD}, Annal. of Phys. \textbf{322} (2007) 1587, arXiv:0704.2798 [hep-ph],\\
L.~J.~Dixon, \emph{Calculating scattering amplitudes efficiently}, arXiv:9601359 [hep-ph],\\
R.~Britto \emph{Loop amplitudes in gauge theories: modern analytic
approaches}, J.\ Phys.\ A \textbf{44}, 454006 (2011), arXiv:1012.4493 v2 [hep-th], \\
Z.~Bern, Yu-tin ~Huang \emph{Basics of Generalized Unitarity},\\
H. Elvang, Yu-tin Huang, \emph{Scattering amplitudes}, arXiv:1308.1697 v1 [hep-th].
\bibitem{Bern}
Z.~Bern, L.~J.~Dixon, D.~C.~Dunbar and D.~A.~Kosower, \emph{One-Loop
n-Point Gauge Theory Amplitudes, Unitarity and Collinear Limits},
Nucl.\ Phys.\  B {\bf 425} (1994) 217, [arXiv:hep-ph/9403226];
\emph{Fusing gauge theory tree amplitudes into loop amplitudes},
Nucl.\ Phys.\  B {\bf 435} (1995) 59, [arXiv:hep-ph/9409265],\\
N. Arkani-Hamed, F. Cachazo, J. Kaplan, \emph{What is the Simplest Quantum Field Theory?},
JHEP \textbf{1009} (2010) 016, arXiv:0808.1446 [hep-th].
\bibitem{UsD} 	N.I. Usyukina, A.I. Davydychev, 
 \emph{Exact results for three and four point ladder diagrams with an arbitrary number of rungs}, 
 Phys.Lett. {\bf B305} (1993) 136.
\bibitem{UsB} V.V. Belokurov, N.I. Usyukina,
 \emph{Calculation Of Ladder Diagrams In Arbitrary Order}, 
 J.Phys. {\bf A16} (1983) 2811. 
\bibitem{Smir}	
V.A. Smirnov,  
 \emph{Analytical result for dimensionally regularized massive on-shell planar double box },
Phys.Lett. {\bf B524} (2002) 129, hep-ph/0111160;
V.A. Smirnov, 
 \emph{Analytical result for dimensionally regularized massless on shell planar triple box }
Phys.Lett. {\bf B567} (2003) 193,  hep-ph/0305142

\bibitem{Arc} 
T.~Dennen, Yu-tin~Huang, \emph{Dual Conformal Properties of
Six-Dimentional Maximal Super Yang-Mills Amplitudes}, JHEP
\textbf{1101} (2011) 140, arXiv:1010.5874 v2 [hep-th],\\
Z.~Bern, J.J.~Carrasco, T.~Dennen, Yu-tin~Huang, and H.~Ita,
\emph{Generalized Unitatity and Six-Dimentional Helicity}, Phys.\
Rev.\ D \textbf{83} 085022 (2011),  arXiv:1010.0494 v2 [hep-th],
S.~Caron-Huot, D.~O'Connel, \emph{Spinor Helicity and Dual Conformal
Symmetry in Ten Dimentions}, JHEP \textbf{1108} (2011) 014,
arXiv:1010.5487 [hep-th],\\
R.~H.~Boels, D.~O'Connel, \emph{Simple superamplitudes in higher
dimensions}, JHEP \textbf{1206} (2012) 163, arXiv:1201.2653 [hep-th].

\bibitem{BKV} L.V. Bork, D.I. Kazakov, D.E. Vlasenko,
 \emph{On the amplitudes in N=(1,1) D=6 SYM},
JHEP {\bf 1311} (2013) 065,  arXiv:1308.0117
\bibitem{SmirBook} V.~A.~Smirnov,
 ``Analytic tools for Feynman integrals,''
 Springer Tracts Mod.\ Phys.\  {\bf 250} (2012) 1.
 \bibitem{Uniq} A.N. Vassiliev, Yu.M.Pismak, Yu.R.Honkonen, Theor.Math.Phys., {\bf 47} (1981) 291;
 N.I.Usyukina, Theor.Math.Phys., {\bf 54} (1983) 124;
 D.I.Kazakov, Theor.Math.Phys., {\bf 58} (1984) 343.
\bibitem{Tausk} C. Anastasiou, J. B. Tausk, M. E. Tejeda-Yeomans, 
{The on-shell massless planar double box diagram with an irreducible numerator},  
Nucl. Phys. Proc. Suppl. {\bf 89} (2000) 262, arXiv:hep-ph/0005328 v1.
\bibitem{Smirnov:2009up}
 A.~V.~Smirnov and V.~A.~Smirnov,
  \emph{On the Resolution of Singularities of Multiple Mellin-Barnes Integrals},
 Eur.\ Phys.\ J.\ C {\bf 62} (2009) 445
 [arXiv:0901.0386 [hep-ph]].
\bibitem{Czakon:2005rk}
 M.~Czakon,
  \emph{Automatized analytic continuation of Mellin-Barnes integrals},
 Comput.\ Phys.\ Commun.\  {\bf 175} (2006) 559
 [hep-ph/0511200];\\
 see also D.A. Kosower, http://projects.hepforge.org/mbtools/
\bibitem{Bogner:2007cr} http://inspirehep.net/record/761982
\bibitem{Hahn:2004fe}http://inspirehep.net/record/647621
\bibitem{Fiesta} A. V. Smirnov, V. A. Smirnov and M. Tentyukov, Comput. Phys. Commun., {\bf 182} (2011) 790,
[arXiv:0912.0158 [hep-ph]].
\end{thebibliography}
\end{document}